# Giant-exchange-driven Vectorial Control of a Minimal Topological Magnet in $Eu_3In_2As_4$

Haonan Chen[1#], Xunkai Duan[2, 3#], Guangyi Wang[4, 5#], Yuhan Du[4, 5], Huayao Li[6], Jiayu Wang[1], Wenbin Wu[4, 5], Zixuan Xu[1], Yingchao Xia[1], Jiaming Gu[1], Pengliang Leng[1], Lin Miao[6], Fengfeng Zhu[7, 8], Xiang Yuan[4, 5*], Tong Zhou[2*], Cheng Zhang[1, 9*]

[1] State Key Laboratory of Surface Physics and Institute for Nanoelectronic Devices and Quantum Computing, Fudan University, Shanghai 200433, China
[2] Ningbo Institute of Digital Twin, Eastern Institute of Technology, Ningbo, Zhejiang 315200, China
[3] School of Physics and Astronomy, Shanghai Jiao Tong University, Shanghai 200240, China
[4] State Key Laboratory of Precision Spectroscopy, East China Normal University, Shanghai 200241, China
[5] School of Physics and Electronic Science, Key Laboratory of Polar Materials and Devices, Ministry of Education, East China Normal University, Shanghai 200241, China
[6] Key Laboratory of Quantum Materials and Devices of Ministry of Education, School of Physics, Southeast University, Nanjing 211189, China
[7] 2020 X-Lab, Shanghai Institute of Microsystem and Information Technology, Chinese Academy of Sciences, Shanghai 200050, China
[8] National Key Laboratory of Materials for Integrated Circuits, Shanghai Institute of Microsystem and Information Technology, Chinese Academy of Sciences, Shanghai 200050, China
[9] Zhangjiang Fudan International Innovation Center, Fudan University, Shanghai 201210, China

[#] These authors contributed equally to this work
[*] Correspondence and requests for materials should be addressed to C. Z. (E-mail: zhangcheng@fudan.edu.cn), T. Z. (E-mail: tzhou@eitech.edu.cn), and X. Y. (E-mail: xyuan@lps.ecnu.edu.cn)

## Abstract

The interplay between magnetism and band topology provides a route to controlling quantum states of matter, yet its realization in materials is often constrained by weak exchange coupling and complex electronic structures. Here, a giant exchange coupling is identified in the newly predicted topological magnet $Eu_3In_2As_4$, giving rise to magnetization-dependent band shifts of up to 300 meV. Together with its intrinsically soft magnetic response, this strong coupling enables systematic tuning of topological phases by both the magnitude and orientation of applied magnetic fields. The magneto-topological phase diagram is mapped out in which an antiferromagnetic topological insulator ground state evolves, under modest fields, into a proposed intermediate 2/3-ferrimagnetic phase, and further into fully polarized ferromagnetic states predicted to host either Weyl or nodal-ring semimetals. Notably, the Weyl phase corresponds to a minimal model hosting a single pair of Weyl nodes. Quantum oscillations, anomalous Hall transport and magneto-infrared spectroscopy consistently reveal exchange-driven band reconstruction across these transitions. Rotation of the magnetization theoretically provides an efficient means to tune the momentum-space positions and separations of the Weyl nodes. These results establish $Eu_3In_2As_4$ as a model system for exploring how strong exchange coupling can be used to control topological band structures with minimal complexity.

## 1. Introduction

The integration of intrinsic magnetism with quantum materials offers a powerful avenue for realizing and dynamically controlling diverse exotic topological phases[1–3]. Central to this synergy is the exchange coupling between localized magnetic moments and itinerant electrons, which breaks time-reversal symmetry (TRS) and profoundly reshapes electronic band structures. In magnetic topological insulators, such coupling opens gaps in surface Dirac cones, enabling axion electrodynamics and the quantum anomalous Hall effect[1, 2, 4–6]. In Weyl semimetals, TRS breaking generates pairs of Weyl nodes with opposite chirality, whose separation in momentum space governs large intrinsic anomalous Hall effect (AHE)[7–10] and anomalous Nernst effect (ANE)[11–14]. The strength and tunability of the exchange coupling thus dictate not only the stability of these phases but also the degree to which they can be manipulated by external fields, a prerequisite for both fundamental exploration and device applications.

Despite significant progress, existing magnetic topological materials face persistent challenges that limit their utility. Many candidates exhibit complex band structures riddled with multiple bands near the Fermi level, obscuring the clean realization of minimal topological models[10, 13, 15–22]. Moreover, the exchange coupling between local moment and itinerant carriers in most systems is weak[15, 23, 24], typically yielding band shifts on the order of a few to tens of meV, rendering topological transitions either inaccessible or requiring extreme conditions. Compounding these issues, low carrier mobility often masks intrinsic topological transport signatures[22, 25–29], while strong magnetic anisotropy restricts field-tunability to narrow parameter windows. These limitations underscore the urgent need for new platforms where strong exchange couplings coexist with simple, well-isolated topological bands and soft magnetic response.

Here, we introduce $Eu_3In_2As_4$ as a model magnetic topological material that overcomes these limitations. We uncover a giant exchange coupling—manifested as magnetization-dependent band shifts up to 300 meV, far exceeding conventional Zeeman scales, which drive pronounced reconstructions of the electronic structure across multiple magnetic phases. $Eu_3In_2As_4$ hosts an antiferromagnetic (AFM) topological insulator ground state below the Néel temperature ($T_N$ = 4.5 K), which evolves under modest magnetic fields into a 2/3-ferrimagnetic intermediate phase, and ultimately into fully polarized ferromagnetic (FM) states predicted to realize ideal Weyl or nodal-ring semimetals. Remarkably, these FM states feature the minimal Weyl model: a single pair of Weyl nodes, free from extraneous band complications. Owing to intrinsic magnetic softness, full polarization is achieved with fields as low as ~1 T along any crystallographic axis, and rotation of the magnetization direction is proposed to efficiently tune the momentum-space positions and separations of the Weyl nodes. Through quantum oscillations, anomalous Hall transport, and magneto-infrared spectroscopy, we provide consistent evidence of exchange-driven band reconstruction and topological phase transitions. Our work establishes $Eu_3In_2As_4$ as a uniquely versatile platform where giant exchange coupling enables unprecedented magnetic control over clean, tunable, and minimal topological band structures.

# 2. Results

### 2.1 Diverse magnetic phases with distinct topological band structures.

$Eu_3In_2As_4$ crystallizes in the orthorhombic space group *Pnnm* (No. 58)[30], adopting a Zintl-phase architecture composed of quasi-one-dimensional [$In_2As_4$] chains extending along the *c*-axis, intercalated by two inequivalent $Eu^{2+}$ sites in the *ab*-plane (Figure 1a). Single crystals were structurally and compositionally characterized via X-ray diffraction (XRD), energy-dispersive X-ray spectroscopy (EDX) and Raman spectrum (see Notes S1 and S2). Energy calculations identify the AFM state with the Néel vector along *a*-axis (AFM*a*) as the magnetic ground state (energy differences listed in Table S3). In this phase, $Eu^{2+}$ moments form a C-type AFM order, where the two inequivalent $Eu^{2+}$ sites constitute two AFM sublattices respectively (Figure 1b).

Magnetic susceptibility $\chi(T)$ measurements (Figure 1c) confirmed the AFM ground state, revealing peaks at $T_N \sim 4.5$ K for all field orientations. Below $T_N$, $\chi(T)$ for $H // a$ sharply drops towards zero, whereas for fields perpendicular to *a*-axis, lower transition peaks and low-temperature upturns are observed. This anisotropy indicates the *a*-axis as the easy magnetization direction. Curie-Weiss fitting (Figure S5) in the paramagnetic (PM) regime yields positive Weiss temperatures for all axes, suggesting short-range FM correlations between $Eu^{2+}$ ions above $T_N$, despite the AFM ground state. Such contrast, previously reported in low-carrier-density $Eu^{2+}$-based antiferromagnets, is attributed to magnetic polaron formation[31–37]. These polarons arise from strong exchange coupling between localized 4*f* moments and itinerant carriers, leading to FM clusters embedded in a PM or AFM matrix[38]. Beyond the positive Weiss temperatures, the low-temperature upturns in $\chi(T)$ for $H // b$ and *c* which signals weak ferromagnetism and the deviation of Curie-Weiss law support this scenario of magnetic polarons (Figure S5).

Figure 1d exhibits the magnetization $M(H)$ curves at 1.8 K. For $H // a$, a low-field plateau below 0.1 T reflects the AFM easy-axis configuration, followed by a sharp magnetization jump resembling a spin-flop transition and an additional anomaly near 0.5 T. In contrast, for $H // b$ or *c*, *M* initially increases linearly at low fields and exhibits slope changes around 0.69 T before gradually approaching saturation. These features are more clearly resolved in $\mathrm{d}M/\mathrm{d}H$ and $\mathrm{d}^2M/\mathrm{d}H^2$ curves. Unlike conventional easy-axis antiferromagnets with weak anisotropy, which typically exhibit a single spin-flop transition, $Eu_3In_2As_4$ displays two successive sets of anomalies in both d$M$/d$H$ and d$^2M$/d$H^2$ curves for $H$ // $a$ (Figure S6), indicating two consecutive field-induced reorientation/polarization processes rather than a single-step transition.

$Eu_3In_2As_4$ contains two crystallographically inequivalent Eu sublattices (Eu1 and Eu2) with multiplicities of four and two per magnetic unit cell, contributing 2/3 and 1/3 of the total saturated moment, respectively. We therefore propose a sequential polarization scenario: starting from the AFM$a$ ground state, the Eu1 sublattice first undergoes a spin-flop transition and evolves into a canted AFM (cAFM-1) state. As the field increases, the Eu1 moments continuously cant towards $a$-axis and reaches full polarization first, yielding an intermediate ferrimagnetic-like state with approximately 2/3 of the saturated magnetization (denoted as 2/3-FiM$a$), while Eu2 sublattices remain AFM-coupled (Figure 1d inset). The Eu2 sublattice subsequently undergoes its own spin-flop and polarization process (cAFM-2), ultimately driving the system into the fully polarized FM$a$ state. This sequential evolution is schematically summarized in Figure S7. Energy calculations support the energetic feasibility of the 2/3-FiMa configuration as an intermediate state between AFMa and FMa states (Table S3). Analogous consecutive magnetization on different Eu-sites was reported in Zintl antiferromagnet $Eu_5In_2As_6$[31].

For $H$ // $b$ and $c$, Eu1 spins continuously rotate from the easy $a$-axis towards the field direction until alignment. The peak features in d$M$/d$H$ again occur when the magnetization reaches approximately 2/3 of the saturated value, consistent with the sequential polarization scenario proposed above. Beyond this field, the remaining Eu2 sublattice progressively aligns with the external field, inducing an increase of $M$ with a more rapid slope until complete spin polarization is achieved. At high fields, $M$ saturates first along the easy $a$-axis, and a small field range (~1 T) is sufficient to achieve full spin polarization for all axes, revealing soft magnetism and weakly-coupled AFM ground state (also evidenced by low $T_N$ and tiny AFM-FM energy difference). The saturated magnetization ($M_{sat}$) reaches ~ 7 $\mu_B$/$Eu^{2+}$, confirming the $Eu^{2+}$-dominated magnetism with half-filled 4$f$ electrons ($S$ = 7/2). $M$($H$) curves measured up to ±7 T with rising temperatures are summarized in Figure S8.

Detailed static magnetization characterizations delineate magnetic phase diagram for $H$ // $a$ in Figure 1g, with the phase boundaries determined from $\chi(T)$, $M(H)$, d$M$/d$H$, and d$^2M$/d$H^2$ curves (see Note S3 for details). Higher fields suppress and shift the $\chi(T)$ peaks to lower temperatures until vanishing (Figure 1e), defining the boundary between the long-range ordered and PM states. At low temperatures, based on the sequential polarization process described above, the ordered region is divided into the AFM$a$, cAFM-1, 2/3-FiM$a$, cAFM-2, and FM$a$ phases (Figure 1f). The characteristic fields $\mu_0H_c$ and $\mu_0H_s$ are extracted from the peaks in d$M$/d$H$ and the minima in d$^2M$/d$H^2$, respectively[39–42], and are used to divide the phase boundary. Here, $\mu_0H_c$ denotes the characteristic field of the spin-flop process, whereas $\mu_0H_s$ marks the onset of the fully polarized ferromagnetic state. Notably, the intermediate 2/3-FiM$a$ phase emerges only at lower temperatures and occupies a finite field window near the $\mu_0H_{s1}$ line. Its microscopic spin configuration is inferred from magnetization and derivative analyses, while a definitive determination would require neutron diffraction or related magnetic-structure probes.

At elevated temperatures, thermal fluctuations suppress the inflection features and broaden the saturation process in $M(H)$, rendering the minima in d$^2M$/d$H^2$ less sharply defined, towards S-shapes in PM regime above $T_N$ (Figure 1f). In this regime, the fields $\mu_0H_p$ for full polarization are derived from the intersections of linear fits to the low-field and high-field regions of $M(H)$, following the procedures used in Refs. [43, 44]. This provides a consistent criterion for tracing the high-temperature boundary of the field-polarized regime. Additional magnetization characterizations are detailed in Note S3.

Density functional theory (DFT) calculations incorporating spin-orbit coupling (SOC) reveal the magnetic-order-driven topological phase evolution in $Eu_3In_2As_4$ (Figure 1h). In the AFM*a* state, SOC reduces the indirect gap near Γ and induces band inversion, giving rise to a topologically nontrivial bandgap. Based on the reported topological surface state calculations, AFM*a* $Eu_3In_2As_4$ is identified as an axion insulator candidate with hybrid-order band topology[45, 46]: magnetism gaps the Dirac surface states of *bc*- and *ac*-plane, whereas the preserved $C_{2z}$T symmetry protects an unpinned Dirac cone on the *ab*-plane, introducing emergent one-dimensional hinge states among the gapped surfaces. For 2/3-FiM*a* state, the bulk bandgap is further reduced; the remaining small gap suggests proximity to the phase boundary between a topological insulator and a semimetal, where the analysis of surface states theoretically consistent with axion-insulator characteristics (details in Figure S12).

Most strikingly, full spin polarization transforms $Eu_3In_2As_4$ into a minimal Weyl semimetal. For FM alignment along *a* (or *b*), band inversion generates a single pair of Weyl nodes along the Γ–X (Γ–Y) path, with no additional trivial bands crossing the Fermi level, fulfilling the criteria for an ideal Weyl semimetal. When magnetization is forced along the *c*-axis, the system instead hosts an ideal nodal-ring within the $k_x$-$k_y$ plane, stabilized by mirror symmetry (band structures for FM*b*/*c* are detailed in Figure S11). This direct linkage between spin orientation and topological band geometry establishes $Eu_3In_2As_4$ as a uniquely programmable platform for exploring clean, exchange-driven Weyl physics.

### 2.2 Magneto-infrared spectroscopy of $Eu_3In_2As_4$.

To experimentally track the evolution of the electronic structure across these distinct magnetic orders, we employed magneto-infrared spectroscopy, which enables direct monitoring of field-dependent optical transitions over a broad energy window. Magneto-reflectivity $R_B$ measurements were performed on the (100) surface of $Eu_3In_2As_4$ at around 6 K, with the field applied along *a*-axis (*c*-axis) for Faraday (Voigt) geometry. The $R_B$ spectra were normalized by the zero-field reflectivity $R_0$, yielding relative magneto-reflectivity $R_B/R_0$, presented as stacked plots (vertically constant offset) in Figure 2a (Faraday geometry) and Figure 2b (Voigt geometry). Spectral peaks were assigned to inter-band optical transitions, and three transitions, labeled $T_\alpha$, $T_\beta$, and $T_\gamma$, are observed in both geometries. Moreover, the first derivatives $\mathrm{d}(R_B/R_0)/\mathrm{d}\omega$ were also plotted in Figure 2c–d to better resolve the peak positions and their field evolution.

As the magnetic field increases, the transition energies display a distinctly non-monotonic evolution (details in Figure S13), characterized by a rapid initial change followed by gradual saturation. This behavior closely tracks the field-dependent magnetization measured at the same temperature. Although experiments were conducted above $T_N$ due to setup limitations, the field-induced fully polarized state is still accessible at high fields, likely facilitated by strong spin fluctuations or short-range FM correlations, consistent with the positive Weiss temperatures. Importantly, the exceptionally large energy shifts—exceeding 300 meV (for $T_\beta$) within 3 T—are far beyond the scale expected for Zeeman splitting[47–52], thereby unequivocally ruling out the Zeeman effect as the underlying mechanism. Instead, these results demonstrate a highly efficient and giant band modulation related to the magnetization.

Similar *M*-dependent optical transitions have been reported in several $Eu^{2+}$-based semiconductors, including EuTe and $EuCd_2X_2$ (X = P, As, Sb)[53–56], where they originate from exchange couplings between localized $Eu^{2+}$ moments and itinerant electrons. We attribute the giant energy shifts observed in $Eu_3In_2As_4$ to the same mechanism.

Microscopically, $Eu^{2+}$ possesses a half-filled $4f^7$ configuration with $S = 7/2$ and $L = 0$, giving rise to a large spin-only magnetic moment of ~7 $\mu_B$. Upon magnetic polarization, these localized moments generate a strong internal exchange field acting on the itinerant carriers. Within the spin-only Heisenberg picture, the exchange coupling mechanism of band splitting can be expressed as[55, 57–59]:

$$H_{\mathrm{ex}} = -2J_{\mathrm{ex}} \sum_i \mathbf{s}_i \cdot \mathbf{S}_i \tag{1}$$

where $\mathbf{s}_i$ and $\mathbf{S}_i$ denote the spins of itinerant electrons and neighboring $Eu^{2+}$ moments, respectively, and exchange constant $J_{\mathrm{ex}}$ characterizes the coupling strength. Under the mean-field approximation, the average local Eu-spin polarization scales with the magnetization $M(T,B)$, according to $\langle \mathbf{S}_i \rangle = S \cdot M(T,B)/M_{\mathrm{sat}}$[55], yielding an exchange-induced band splitting energy:

$$E_{\mathrm{ex}} = -2J_{\mathrm{ex}} \cdot \frac{1}{2} \cdot S \cdot M(T,B)/M_{\mathrm{sat}} = -J_{\mathrm{ex}} S M(T,B)/M_{\mathrm{sat}} \tag{2}$$

The corresponding band shift energy of a single spin branch is $E_{\mathrm{ex}}/2$. Because the observed infrared transition involves two different electronic bands, the measured transition-energy shift reflects the combined exchange responses of both the initial and final states. We therefore introduce an effective exchange constant $J_{\mathrm{eff}}$, representing the net exchange contributions of the two participating bands. The field dependence of the transition energy can then be written as[53, 54]:

$$\Delta E = E(B) - E_0 = -\frac{1}{2} J_{\mathrm{eff}} S M(T,B)/M_{\mathrm{sat}} \tag{3}$$

where $E(B)$ and $E_0$ are the transition energies at finite and zero field, respectively. This relation predicts a linear scaling of transition energies with magnetization rather than with magnetic field itself. The extracted transition energies are plotted in Figure 2e–f for both geometries, where the field-dependent energy shifts of $T_\alpha$, $T_\beta$, and $T_\gamma$ are well captured by linear scaling with magnetization (solid shaded curves). Notably, the near-identical extrapolation to zero field of those three transitions in both geometries reveals a common band origin. Slight differences in scaling parameters between Faraday (Figure 2e) and Voigt (Figure 2f) geometries indicate minor exchange anisotropy, consistent with the soft and nearly isotropic magnetism of FM-$Eu_3In_2As_4$.

Additionally, a clear decrease was observed in relative magneto-reflectivity within the 600–800 $cm^{-1}$ range, likely associated with a redshift of the plasma frequency (Figure S14). Figure 2g illustrates a simplified model of the optical transitions. At zero field, spin-degenerate bands (in black) give rise to only two transitions: $T_\alpha$ and $T_\beta$ originate from the lower band (LB) to the middle band (MB), while $T_\gamma$ likely corresponds to a transition from the LB to upper band (UB). With increasing field, exchange-induced spin splitting shifts the spin-up bands (red) to lower energy and the spin-down bands (blue) to higher energy. Owing to the difference in the exchange splitting scales among the relevant bands, the transition energies respond differently to magnetization: $T_\alpha$ and $T_\gamma$ redshift with increasing magnetization, whereas $T_\beta$ blueshifts. Within this framework, the exchange splitting parameters can be quantitatively extracted, yielding a giant band splitting approaching ~500 meV for MB (at 12 T) (see Tables S4–S5).

Other transitions anticipated within this level scheme are not experimentally resolved (see Figure S15), as a natural consequence of phase-space constraints and instrumental boundaries. In particular, transitions originating from $LB_\downarrow$ are suppressed by Pauli blocking, while higher-energy channels such as $LB_\uparrow \rightarrow UB_\downarrow$ lie beyond the detection window of the InSb detector. Although the schematic model neglects SOC and antiferromagnetic exchange, these effects are expected to renormal-ize the detailed level structure without altering the overarching physical picture. Crucially, the extracted exchange energy scale is sufficiently large to drive gap closure and band inversion in the calculated fully spin-polarized phase, thereby providing a direct experimental basis for exchange-driven low-energy band reconstruction and topological phase transitions.

### 2.3 Anisotropic magneto-transport of $Eu_3In_2As_4$.

To access the low-energy electronic states that govern charge transport across different magnetic orders, we therefore carried out anisotropic magneto-transport measurements on $Eu_3In_2As_4$. $Eu_3In_2As_4$ exhibits metallic behavior upon cooling, with a resistivity anomaly near $T_N$ (see Figure S16). Figure 3a displays the field-dependent resistivity $\rho_{xx}(H)$ measured up to 2 T, with current applied along the *c*-axis and field rotated within *ab*-plane by angle $\theta$ (inset, Figure 3a). The in-plane magnetoresistance (MR) anisotropy is further visualized in the angular-resolved $\rho_{xx}(\theta,H)$ mapping (Figure 3b).

For all field orientations, $\rho_{xx}(H)$ initially decreases, reaches a minimum at a critical field, and subsequently increases as field-forced FM states develop. Band calculations reveal a small bandgap in AFM*a* ground state, which continuously closes upon transitioning to gapless FM semimetals, likely contributing to the initial negative magnetoresistance (NMR). Alternatively, the initial NMR may originate from field-driven percolation transition of magnetic polarons, leading to carrier delocalization and thus reduced resistivity[35]. With further increasing the field, the MR changes sign and becomes positive before complete spin-polarization. As shown in Figure 3c, where we compare $\rho_{xx}(H)$ and $M(H)$ along three crystalline axes, the resistivity features closely track the magnetization steps (indicated by gray dashed lines), demonstrating that the transport response is strongly affected by the field-induced evolution of the magnetic state.

Upon entering the FM states, $\rho_{xx}(H)$ increases monotonically without additional inflection points (see Figure 4a). The high-field positive, non-saturating MR is likely dominated by orbital magnetoresistance from high-mobility carriers and is consistent with the calculated topological semimetallic state in FM phases. To further quantify the transport anisotropy, the angular-dependent magnetoresistance (AMR), defined as $\mathrm{AMR} = \frac{[\rho_{xx}(\theta,H) - \rho_{xx}(0°,H)]}{\rho_{xx}(0°,H)} \times 100\%$, is shown in Figure 3d. At 2 K and selected fields, AMR exhibits a clear twofold symmetry with maxima along the *a*-axis and minima for *b*-axis. As the field increases, the AMR peak at 90° (*H* // *a*) initially grows rapidly up to 0.3 T, then decreases continuously towards zero at higher fields, indicating diminishing low-field anisotropy as FM state becomes nearly isotropic. Figure 3e summarizes the field-dependent AMR at 90° with increasing temperature. The pronounced non-monotonic field dependence at 2 K (rapid increase to a maximum followed by continuous decrease) is gradually suppressed at higher temperatures and vanishes above $T_N$, yielding negligible AMR at 10 K. Insets of Figure 3e compare the AMR($H$,$T$) map at 90° with the magnetization anisotropy between *a*- and *b*-axis, $M_a(H,T)-M_b(H,T)$, revealing similar non-monotonic trends below $T_N$ and underscoring the intimate correlation between AMR and anisotropic magnetic order within low-field range. By contrast, field rotation in the *bc*-plane yields relatively small AMR (see Figure S25), consistent with the nearly isotropic $M(H)$ response within *bc*-plane.

Unlike the conventional Fermi-surface-anisotropy-induced AMR observed in nonmagnetic materials, whose peak amplitude typically increases monotonically with field[60–62], the non-monotonic field-dependence of AMR in $Eu_3In_2As_4$ resembles that reported in narrow-gap AFM semi-conductors[63–65], where exchange-driven band reconstruction plays a dominant role. In such systems, magnetic anisotropy controls the spin configuration, while magnetization process along specific field directions reduces or even closes the bandgap.

Following this framework, we attribute the AMR to the combined effects of anisotropic magnetic scattering and exchange-driven band reconstruction. In $Eu_3In_2As_4$, first-principles calculations predict the electronic structure evolves continuously from a gapped AFM state to a gapless FM state with topological band crossings. Because the exchange-induced band shift is proportional to $M(H)$, the bandgap evolution is expected to follow the magnetic polarization process. Consequently, the close correspondence between the AMR and the magnetization anisotropy provides transport evidence that the low-energy electronic structure is continuously reconstructed by the exchange coupling between localized $Eu^{2+}$ moments and itinerant carriers.

**2.4 Quantum oscillations in FM phase $Eu_3In_2As_4$.**

Shubnikov-de Haas (SdH) oscillations were analyzed to further elucidate the Fermi surface topology of $Eu_3In_2As_4$. As shown in Figure 4a, pronounced quantum oscillations emerge in $\rho_{xx}(H)$ above ~8 T, a field sufficient to fully saturate the spin polarization and stabilize the FM band structure through exchange coupling. In this regime, the Landau-level spacing exceeds disorder broadening, enabling clear observation of SdH oscillations. Upon rotating the field from $b$- to $a$-axis (with angle $\theta$), SdH oscillations persist in $\rho_{xx}(H)$. Meanwhile, once $Eu_3In_2As_4$ enters FM states, $\rho_{xx}(H)$ exhibits a quasi-linear, non-saturating increase, reminiscent of transport observations in Weyl systems[18, 66–71].

After subtracting non-oscillatory backgrounds via polynomial fitting, the oscillatory component $\Delta\rho_{xx}$ is plotted against $1/\mu_0 H$ (Figure 4b; curves are vertically offset by 1 μΩ·cm). While oscillation amplitude remains largely unchanged, systematic phase evolutions occur with field direction. Owing to the limited oscillation window (~4 T), the oscillation frequency $F$ was extracted from Landau fan diagrams (Figure S17a) rather than fast Fourier transform (FFT) analysis for its insufficient frequency resolution. A single dominant frequency ($F$ ~124 T) is resolved for all directions, showing weak angular dependence (Figure 4c). Such behavior indicates a relatively simple and nearly isotropic three-dimensional (3D) Fermi surface in FM $Eu_3In_2As_4$, consistent with the simple-band-structure picture predicted by calculations. Through Onsager's relation, $F = (\hbar/2\pi e)S_F$, the frequency corresponds to the extremal cross-sectional area $S_F$ of the Fermi pocket perpendicular to the field, yielding $S_F = 1.18\ nm^{-2}$. Based on this $S_F$, we compute the 3D Fermi surface for different magnetization directions (Figure 4d), which remains close to an isotropic ellipsoid under magnetization rotation, consistent with the nearly angle-independent single-frequency SdH response. The experimental Fermi level is determined by matching the extracted $S_F$ to the calculated Fermi surface cross-sectional area, yielding a value approximately 0.155 eV above the DFT Fermi energy, placing it above the Lifshitz transition. Besides, the large electron pocket aligns with the high electron density ($n_e = 2.1\times10^{19}\ cm^{-3}$ and carrier mobility $\mu_e = 1015.1\ cm^2/V\cdot s$) derived from linear fitting of Hall signals (inset, Figure 4a), confirming that transport is dominated by a single high-mobility electron pocket.

Beyond the oscillation frequency, the SdH phase shift $\Phi$ is obtained from Landau-fan $y$-intercept, providing a complementary probe of the Berry-phase ($\phi_B$) and band topology. Trivial parabolic bands yield $\phi_B = 0$, whereas linear band dispersions typically approach $\pi$-Berry phase. The extracted $\Phi$ values across field rotation are consistent with a $\pi$-Berry phase, supporting the presence of topologically nontrivial bands in the FM-$Eu_3In_2As_4$ under rotating magnetization (see Figure S17b).

In the fully polarized FM regime, $Eu_3In_2As_4$ behaves as a soft magnet whose spin moments continuously follow the field direction, rendering its band topology intrinsically angle-tunable. First-principles calculations predict a minimal Weyl platform: FM$a$ (FM$b$) state supports an ideal Weyl semimetal with a single Weyl pair along Γ–X (Γ–Y). Upon rotating the magnetization within the $ab$-plane, canted-spin calculations show that the original Γ–Y Weyl crossing in FM$b$ progressively gaps out while a gap simultaneously closes along Γ–X, culminating in a new Weyl pair in FM$a$ limit (Figure S18). Throughout this rotation, the system remains Weyl semimetallic. To visualize the Weyl cones, we calculated 3D band structures for a series of canting angles (Figure S19); a representative result at $\theta = 45°$ is shown in Figure 4e, revealing a single Weyl-node pair. Figure 4f summarizes the calculated node trajectories, showing migration from high-symmetry lines to general $k$-points in the $k_x$-$k_y$ plane, providing a direct handle to steer their momentum-space positions. As an independent verification, the Weyl-node search using WannierTools likewise identifies one Weyl pair residing in the $k_z = 0$ plane (Figure S19). By contrast, spin-rotation in the $bc$-plane largely preserves the ideal Weyl points along Γ–Y, until the FM$c$ limit, where a nodal-ring semimetal forms in the $k_x$-$k_y$ plane (Figure S18). Overall, $Eu_3In_2As_4$ enables rare vectorial tuning of a minimal Weyl phase via magnetization reorientation.

To further probe the topological electronic structure in the field-polarized state, we examined the anomalous Hall effect (AHE), which provides a sensitive transport probe of the Berry-curvature distribution and is quantified by anomalous Hall conductivity, $\sigma_{\mathrm{AHE}}$. For an ideal type-I FM Weyl semimetal with the Fermi level located below the Lifshitz transition, the intrinsic $\sigma_{\mathrm{AHE}}$ exhibits Fermi-energy-independence and is only determined by the Weyl point separation $\Delta k$ in momentum-space, $\sigma_{\mathrm{AHE}} = (e^2/2\pi h)\Delta k$[7]. In contrast, when the Fermi level is shifted away from the Weyl nodes or the Weyl cones become highly tilted (type-II), additional occupied-state and free-carrier contributions can introduce $E_{\mathrm{F}}$ dependence[7, 8, 72, 73]. We extracted $\rho_{\mathrm{AHE}}$ by subtracting the linear ordinary Hall background from $\rho_{yx}$, followed by conversion to conductivity $\sigma_{\mathrm{AHE}}$. Figure 4g shows $\sigma_{\mathrm{AHE}}(H)$ with rotating fields. For all curves, $\sigma_{\mathrm{AHE}}$ remains negligible in AFM state, increases sharply across magnetic transitions towards FM order, and eventually saturates. The angular dependence of $\sigma_{\mathrm{AHE}}$ at 7 T (selected to avoid interference from SdH oscillations) is well-fitted by a cosine function (Figure 4h), primarily reflecting the geometric projection with field rotation. While band calculations indicate that magnetization rotation modifies Weyl point positions and separations, the absence of pronounced anomalies in angular-dependent $\sigma_{\mathrm{AHE}}$ suggests that the measured AHE is less sensitive to subtle Weyl cone evolution, likely due to the high electron density and the large Fermi pocket being away from the Weyl points.

To directly evaluate the Berry-curvature contribution at the experimentally relevant conditions, we calculated the intrinsic $\sigma_{\mathrm{AHE}}$ over a broad energy range in FM-$Eu_3In_2As_4$. As shown in Figure S21, the intrinsic $\sigma_{\mathrm{AHE}}$ is large near the Weyl nodes but remains finite at higher energies. At the experimentally determined Fermi level, the calculated $\sigma_{ac,\mathrm{AHE}}$ for $H$ // $b$ is approximately -10 $\Omega^{-1}\cdot\mathrm{cm}^{-1}$, which has the same sign and a comparable magnitude as the experimental value (-28.62 $\Omega^{-1}\cdot\mathrm{cm}^{-1}$). This agreement indicates that Berry curvature associated with the calculated topological band structure contributes to the transport response, while the remaining quantitative discrepancy most likely originates from extrinsic contributions, such as skew-scattering and side-jump mechanisms. Scaling law analysis between the $\sigma_{\mathrm{AHE}}$ and $\sigma_{xx}$ is also performed to distinguish intrinsic and extrinsic contributions (Supporting Information, Note S8).

The evolution of SdH oscillations with field rotated in the $bc$-plane is shown in Figure S25. Moreover, temperature-dependent SdH oscillations (Figure S26) show that, while the oscillation amplitude is thermally damped, oscillations remain detectable even above $T_N$, consistent with FM correlations in PM regime that facilitates full spin polarization at high fields. More importantly, the extrema positions and oscillation frequency shift systematically with temperature. As shown in Figure S26c, the SdH frequency increases by ~15% from 2 K to 20 K, which is anomalous for a rigid-band semimetal where the frequency is typically temperature-independent.

This frequency evolution clearly indicates a temperature-driven Fermi pocket reconstruction and can be naturally explained by the exchange-coupled nature of the $Eu_3In_2As_4$ band structure. The FM band structure is stabilized by exchange coupling to $Eu^{2+}$ moments, thus, the thermal suppression of $M_{sat}$ reduces the exchange-induced band splitting and modifies the extremal orbit area. A quantitative comparison between temperature evolution of the SdH frequency and the $M_{sat}$ shows that the relative frequency shift, $F(T) - F(2\ \mathrm{K})$, closely scales with the change in saturation magnetization, $M_{\mathrm{sat}}(T) - M_{\mathrm{sat}}(2\ \mathrm{K})$ (Figure S26e). This strong correspondence suggests that the $Eu^{2+}$ exchange field plays a dominant role in the Fermi-surface reconstruction. The qualitative derivation is detailed in Supporting Information Note S9.

Together with the magneto-infrared and AMR results presented above, these transport measurements consistently support the crucial role of exchange coupling in governing the electronic structure of $Eu_3In_2As_4$.

## 3. Discussion

In contrast to the inter-site exchange interactions between localized magnetic moments that stabilize long-range magnetic order[74], we focus here on the exchange coupling between localized spins and itinerant carriers. Such spin-carrier coupling plays a central role in the emergence of topological electronic phases, including quantum anomalous Hall states[59], one-dimensional Weyl modes[75], and three-dimensional Van Hove singularities[55]. In $Eu_3In_2As_4$, we observe an unusually large exchange-driven band modulation, with characteristic energy shifts reaching ~300 meV. This energy scale is distinct from the bare exchange splitting commonly encountered in conventional ferromagnetic metals. Instead, it reflects a giant effective reconstruction of low-energy itinerant bands induced by exchange coupling in a weakly screened, low-carrier-density semimetal. Figure 5a compares the characteristic energy scales and effective average field slopes ($\Delta E/\Delta B$) across different band modulation mechanisms, compiled from magneto-optics, photoluminescence, or scanning tunneling microscopy studies. For saturating systems such as $Eu_3In_2As_4$, values correspond to saturation points; whereas for non-saturating cases, maximum experimental energy shifts are used. Zeeman effect yields a weaker linear $B$-dependence of band splitting, observable only under high fields or in systems with large $g$-factor. Landau level transitions display non-saturating, transition-index-dependent energy shifts. In contrast, exchange-driven band modulation exhibit both large energy scales and steep sub-saturation field dependence, particularly evident in $Eu_3In_2As_4$.

The origin of this strong exchange coupling in $Eu_3In_2As_4$ lies in the unique electronic configuration of $Eu^{2+}$. The half-filled $4f^7$ shell enforces, via Hund's rules, a maximal spin angular momentum ($S = 7/2$) and zero orbital angular momentum ($L = 0$). This yields a large and purely spin-derived local magnetic moment of 7 $\mu_B$ as a giant internal exchange field, exceeding that typical of $d$-orbital systems. Crucially, the absence of $L$ avoids higher-rank multipolar interactions associated with total angular momentum $J$, which are common in other $4f$-magnets[58, 76]. This characteristic simplifies the exchange coupling mechanism to spin-only Heisenberg-form band splitting as $H_{\text{ex}} = -2J_{\text{ex}} \sum_i \mathbf{s}_i \cdot \mathbf{S}_i$[55, 57–59], as the Equation (1) above. This form establishes a direct connection between the exchange-induced band splitting or shift and the macroscopic magnetization.

Furthermore, $L = 0$ renders $Eu^{2+}$ insensitive to crystal electric fields, leading to isotropic and soft magnetism in $Eu_3In_2As_4$, enabling Eu moment saturation and reorientation under fields below ~1 T along different axes, essential for efficient vectorial control of the band topology: the external field rotates the Eu-spin polarization, while the exchange coupling transfers this rotation to the itinerant band structure. This sharply contrasts with $4f$ ions possessing nonzero $L$, where significant crystal-field splitting introduces pronounced magnetic anisotropy (*e.g.*, $Tb^{3+}$ in $TbTi_3Bi_4$[77, 78], $Dy^{3+}$ in quasi-Ising $DyF_3$[79]), suppressing local moment along specific directions. It also differs from typical $3d$ magnets, where stronger anisotropy, orbital-hybridization-suppressed local moments, and defect-sensitive magnetic order make field control more difficult, such as $MnBi_2Te_4$[80–82].

In addition, as a Zintl phase[83–86], $Eu_3In_2As_4$ features $Eu^{2+}$ cations occupying the voids of covalently bonded In-As anionic framework with well-defined valence, placing the magnetic moments in close proximity to the conductive channels and promoting appreciable exchange coupling with sizable $J_{\text{ex}}$. Meanwhile, the In-As framework yields a narrow bandgap with the conduction- and valence-band edges (In-$s$ and As-$p$[45]) near $E_F$. The characteristics of low carrier density, narrow band gap, and sparse bands near $E_F$ reduce metallic screening, allowing exchange-driven band shifts to play an outsized role in reshaping the topological properties. Together with the large isotropic $Eu^{2+}$ moment, strong $4f$ localization, and the absence of crystal-field/hybridization-driven moment suppression, this simple band structure accounts for the giant exchange splitting in $Eu_3In_2As_4$, as illustrated in Figure 5b.

This capability is particularly impactful for realizing the ideal Weyl state, featuring a single pair of Weyl points, which remains experimentally rare. Most magnetic Weyl semimetals possess complex band structures with multiple Weyl cones or coexisting trivial Fermi pockets (*e.g.*, $Co_3Sn_2S_2$[10, 25]), obscuring intrinsic Weyl physics and often accompanied by reduced carrier mobility. Although ideal Weyl phases have been proposed in several materials[4, 87–90], experimental realizations remain scarce such as FM-$Mn(Bi_{1-x}Sb_x)_2Te_4$[72, 73] and $(Cr,Bi)_2Te_3$[90]. In $Mn(Bi_{1-x}Sb_x)_2Te_4$, moderate exchange coupling and strong magnetic anisotropy intrinsic to $3d$ magnetism require large fields to achieve FM polarization (~8 T along the $c$-axis and ~11 T in the $ab$-plane[80]) and antisite defects further suppress both magnetic and topological properties, even driving a degradation to trivial ferromagnetic insulators at high defect densities[42]. Other candidates, such as $K_2Mn_3(AsO_4)_3$ and $EuCd_2As_2$, are challenged by recently reported semiconducting or insulating behavior[53, 91–93].

Beyond its nontrivial band topology, DFT calculations further reveal the altermagnetism in AFM-$Eu_3In_2As_4$, enriching the material's phase landscape. Distinct spin-split bands emerge along Γ–S and O–B paths, originating from the crystal symmetry associated with oppositely spin-polarized sublattices (see Figure S29).

# 4. Conclusions

In summary, $Eu_3In_2As_4$ emerges as a uniquely efficient and versatile platform for exploring exchange-driven band engineering and magnetic topological phases, combining large band tunability, high carrier mobility, and exceptional magnetic controllability. The exchange-driven band modulation is directly visualized by large, magnetization-dependent optical-transition energy shifts, reaching ~300 meV within 3 T. Strong coupling to $Eu^{2+}$ 4*f* moments yields a rich phase diagram, spanning an AFM*a* topological insulator ground state, magnetization-direction-tunable ideal FM Weyl semimetals, an ideal nodal-ring semimetal in FM*c* state, and a previously unexplored 2/3-FiM*a* phase. Although the specific topological assignments are based primarily on first-principles calculations, magnetization, AMR, SdH oscillations, AHE measurements and magneto-infrared spectroscopy collectively support exchange-driven band reconstruction across these magnetic phases. Direct spectroscopic observation of the Weyl cones and transport experiments closer to charge neutrality remain important goals for future work, calling for the Fermi-level tuning via optimized growth/chemical substitution, and on low-dimensional devices. Notably, $Eu_3In_2As_4$ nanowires have been synthesized via topotaxial mutual-exchange method[46], underscoring the device feasibility for gate-tunable platforms to access transport evidence of exotic boundary states.

# 5. Methods

### 5.1 Crystal growth and characterization.

High-quality $Eu_3In_2As_4$ crystals were synthesized by flux method[94]. Starting materials of high-purity Eu, In, As and Sn (molar ratio 3: 15: 4.2: 15) were mixed into an alumina crucible fitted with an alumina filter within an Ar-filled glovebox ($H_2O$ and $O_2$ < 0.01 ppm). The crucible was flame-sealed under vacuum in a quartz ampule. Heating was performed in a box furnace (KSL-1200X-J, Kejing) with the following temperature profile: ramp to 1000 °C in 24 h, hold for 48 h to ensure homogenization and reaction, slowly cool to 800 °C at 1 °C/h, and dwell for 24 h. Crystals were subsequently separated from the excess In-Sn flux via centrifugation. This method yielded $Eu_3In_2As_4$ crystals reaching several millimeters in length. Crystal structure and surface orientation were characterized by XRD using a Bruker D8 Discover diffractometer with Cu $K_{\alpha 1}$ radiation ($\lambda$ = 1.54056 Å), achieved using a 2-bounce crystal monochromator to eliminate $K_{\alpha 2}$ component. Elemental composition was verified by EDX (Oxford X-Max). Room-temperature Raman spectrum was measured using a home-built system with 632.8 nm He-Ne laser excitation to detect Raman-active phonon modes.

### 5.2 Magnetization and electrical transport measurements.

Magnetization measurements were performed using VSM mode in a magnetic property measurement system (MPMS3, Quantum Design, ±7 T) with standard quartz sample holder. For electrical transport measurements, needle-shaped samples were firstly polished to eliminate the residual In-Sn flux on the surface which are superconducting at low temperatures. Aluminum wires were bonded to the surface using silver paste (DuPont, 4929N) to form a standard six-wire Hall configuration. The magneto-transport measurements were performed using a Cryogenic superconducting magnet (1.6 ~ 300 K, ±12 T). A voltage-controlled current source (Stanford Research System, CS580) was used to provide an alternating current at 17.777 Hz, and the voltage signals were recorded by several lock-in amplifiers (Stanford Research System, SR865).

**5.3 Magneto-infrared measurement.**

Magneto-infrared spectroscopy was conducted at East China Normal University (ECNU, Shanghai) based on home-built system consists of a FTIR spectrometer and a 12 T closed-cycle superconducting magnet[95]. The collimated infrared beam from a globar/tungsten light source was directed into the variable temperature insert (VTI) of the magnet and focused onto the (100) surface of the $Eu_3In_2As_4$ crystal using an on-axis parabolic mirror which was cooled to 6 K using a small amount of helium exchange gas. The reflected infrared beam was guided out of the VTI through a gold-coated brass light pipe and collected by a liquid-nitrogen-cooled Indium Antimonide detector in a detector chamber away from the magnet.

**5.4 Theoretical calculations.**

The geometry and electronic structures of the studied were calculated by using the projector augmented-wave (PAW)[96] formalism based on density functional theory (DFT), as implemented in the Vienna Ab initio Simulation Package (VASP)[97]. The Perdew-Burke-Ernzerhof generalized gradient approximation (GGA-PBE)[98] was employed to describe the exchange and correlation functional. The kinetic energy cutoff of the plane-wave basis was set to 550 eV, with a total energy convergence criterion of $10^{-6}$ eV. The Brillouin zone integration was performed by using 8 × 4 × 12 Γ-centered k point mesh. All atoms in the unit cell were allowed to relax until the Hellmann-Feynman force on each atom was less than 0.001 eV/Å. To treat the correlation effect of localized 4*f* electrons of Eu, the DFT+*U* method[99] was employed with $U_{\mathrm{eff}} = 7$ eV. The phonon band structure was calculated using the density functional perturbation theory (DFPT)[100]. To investigate the topological properties, the maximally localized Wannier functions (MLWFs) were constructed by using the WANNIER90 package[101]. The edge states, three-dimensional bands, the search for Weyl points and the energy-dependent anomalous Hall conductivity were calculated using the WANNIERTOOLS package[102].

## Author Contributions

C.Z. conceived the idea and supervised the overall research. H.C. synthesized $Eu_3In_2As_4$ single crystals and performed structural and compositional characterizations. H.C. conducted magnetization measurements and analyzed the data with input from F.Z. X.D. and T.Z. carried out the first-principles calculations. G.W. performed magneto-infrared spectroscopy with assistant from Y.D., W.W., and X.Y. H.C. performed electrical transport measurements with the help of J.W., Z.X., Y.X., J.G., and P.L. G.W. and Y.D. performed Raman measurements on $Eu_3In_2As_4$. H.C. and C.Z. analyzed and interpreted the results with contributions from H.L., P.L., L.M., F.Z., T.Z. and X.Y. H.C. and C.Z. wrote the paper with assistance from all other co-authors.

## Acknowledgments

We acknowledge the support by the National Key R&D Program of China (Grants No. 2022YFA1405700, 2023YFA1407500), the National Natural Science Foundation of China (Grants No. 92365104, 12174069, U24A2012, and 12474155), Shanghai Pilot Program for Basic Research-Fudan University 21TQ1400100 (25TQ001), Innovation Project for Integration of Science and Education from Shanghai Institute of Technical Physics, CAS (SITPKJRH-2025-02), Shanghai Qi-Yuan Innovation Foundation, Scientific Research Innovation Capability Support Project for Young Faculty (Grant No. ZYGXQNJSKYCXNLZCXM-M11), Shanghai Pilot Program for Basic Re-search (Grant No. TQ20240203), Shanghai Rising-Star Program (Grant No. 24QA2702200), Shuguang Program (Grant No. 24SG29), and the Zhejiang Provincial Natural Science Foundation of China (LR25A040001).

## Conflicts of Interest

The authors declare no competing interests.

## Data Availability Statement

All raw and derived data supporting the findings in the study are available from the corresponding author upon reasonable request.

## References

1. B. A. Bernevig, C. Felser, H. Beidenkopf, "Progress and prospects in magnetic topological materials," *Nature* 603 (2022): 41–51, https://doi.org/10.1038/s41586-021-04105-x.
2. Y. Tokura, K. Yasuda, A. Tsukazaki, "Magnetic topological insulators," *Nature Reviews Physics* 1 (2019): 126–143, https://doi.org/10.1038/s42254-018-0011-5.
3. J. Zou, Z. He, G. Xu, "The study of magnetic topological semimetals by first principles calculations," *npj Computational Materials* 5 (2019): 96, https://doi.org/10.1038/s41524-019-0237-5.
4. D. Zhang, M. Shi, T. Zhu, D. Xing, H. Zhang, J. Wang, "Topological Axion States in the Magnetic Insulator $MnBi_2Te_4$ with the Quantized Magnetoelectric Effect," *Physical Review Letters* 122 (2019): 206401, https://doi.org/10.1103/PhysRevLett.122.206401.
5. C.-Z. Chang, J. Zhang, X. Feng, et al., "Experimental Observation of the Quantum Anomalous Hall Effect in a Magnetic Topological Insulator," *Science* 340 (2013): 167–170, https://doi.org/10.1126/science.1234414.
6. Y. Deng, Y. Yu, M. Z. Shi, et al., "Quantum anomalous Hall effect in intrinsic magnetic topological insulator $MnBi_2Te_4$," *Science* 367 (2020): 895–900, https://doi.org/10.1126/science.aax8156.
7. A. A. Burkov, "Anomalous Hall Effect in Weyl Metals," *Physical Review Letters* 113 (2014): 187202, https://doi.org/10.1103/PhysRevLett.113.187202.
8. A. A. Zyuzin, R. P. Tiwari, "Intrinsic anomalous Hall effect in type-II Weyl semimetals," *JETP Letters* 103 (2016): 717–722, https://doi.org/10.1134/S002136401611014X.
9. P. Li, J. Koo, W. Ning, et al., "Giant room temperature anomalous Hall effect and tunable topology in a ferromagnetic topological semimetal $Co_2MnAl$," *Nature Communications* 11 (2020): 3476, https://doi.org/10.1038/s41467-020-17174-9.
10. E. Liu, Y. Sun, N. Kumar, et al., "Giant anomalous Hall effect in a ferromagnetic kagome-lattice semimetal," *Nature Physics* 14 (2018): 1125–1131, https://doi.org/10.1038/s41567-018-0234-5.
11. J. Noky, J. Gayles, C. Felser, Y. Sun, "Strong anomalous Nernst effect in collinear magnetic Weyl semimetals without net magnetic moments," *Physical Review B* 97 (2018): 220405(R), https://doi.org/10.1103/PhysRevB.97.220405.
12. M. Ikhlas, T. Tomita, T. Koretsune, et al., "Large anomalous Nernst effect at room temperature in a chiral antiferromagnet," *Nature Physics* 13 (2017): 1085–1090, https://doi.org/10.1038/nphys4181.

13. A. Sakai, Y. P. Mizuta, A. A. Nugroho, et al., "Giant anomalous Nernst effect and quantum-critical scaling in a ferromagnetic semimetal," *Nature Physics* 14 (2018): 1119–1124, https://doi.org/10.1038/s41567-018-0225-6.
14. Y. Pan, C. Le, B. He, et al., "Giant anomalous Nernst signal in the antiferromagnet $YbMnBi_2$," *Nature Materials* 21 (2022): 203–209, https://doi.org/10.1038/s41563-021-01149-2.
15. E. Cheng, L. Yan, X. Shi, et al., "Tunable positions of Weyl nodes via magnetism and pressure in the ferromagnetic Weyl semimetal CeAlSi," *Nature Communications* 15 (2024): 1467, https://doi.org/10.1038/s41467-024-45658-5.
16. K. Ueda, T. Yu, M. Hirayama, et al., "Colossal negative magnetoresistance in field-induced Weyl semimetal of magnetic half-Heusler compound," *Nature Communications* 14 (2023): 6339, https://doi.org/10.1038/s41467-023-41982-4.
17. C. Li, J. Zhang, Y. Wang, et al., "Emergence of Weyl fermions by ferrimagnetism in a noncentrosymmetric magnetic Weyl semimetal," *Nature Communications* 14 (2023): 7185, https://doi.org/10.1038/s41467-023-42996-8.
18. S. Lei, K. Allen, J. Huang, et al., "Weyl nodal ring states and Landau quantization with very large magnetoresistance in square-net magnet $EuGa_4$," *Nature Communications* 14 (2023): 5812, https://doi.org/10.1038/s41467-023-40767-z.
19. T. Suzuki, R. Chisnell, A. Devarakonda, et al., "Large anomalous Hall effect in a half-Heusler antiferromagnet," *Nature Physics* 12 (2016): 1119–1123, https://doi.org/10.1038/nphys3831.
20. S. Nakatsuji, N. Kiyohara, T. Higo, "Large anomalous Hall effect in a non-collinear antiferromagnet at room temperature," *Nature* 527 (2015): 212–215, https://doi.org/10.1038/nature15723.
21. K. Kim, J. Seo, E. Lee, et al., "Large anomalous Hall current induced by topological nodal lines in a ferromagnetic van der Waals semimetal," *Nature Materials* 17 (2018): 794–799, https://doi.org/10.1038/s41563-018-0132-3.
22. I. Belopolski, K. Manna, D. S. Sanchez, et al., "Discovery of topological Weyl fermion lines and drumhead surface states in a room temperature magnet," *Science* 365 (2019): 1278–1281, https://doi.org/10.1126/science.aav2327.
23. S.-K. Bac, F. Le Mardelé, J. Wang, et al., "Probing Berry Curvature in Magnetic Topological Insulators through Resonant Infrared Magnetic Circular Dichroism," *Physical Review Letters* 134 (2025): 016601, https://doi.org/10.1103/PhysRevLett.134.016601.
24. R. Lou, A. Fedorov, L. Zhao, A. Yaresko, B. Büchner, S. Borisenko, "Signature of weakly coupled f electrons and conduction electrons in magnetic Weyl semimetal candidates PrAlSi and SmAlSi," *Physical Review B* 107 (2023): 035158, https://doi.org/10.1103/PhysRevB.107.035158.
25. J. Shen, Q. Yao, Q. Zeng, et al., "Local Disorder-Induced Elevation of Intrinsic Anomalous Hall Conductance in an Electron-Doped Magnetic Weyl Semimetal," *Physical Review Letters* 125 (2020): 086602, https://doi.org/10.1103/PhysRevLett.125.086602.
26. J. Park, G. Lee, F. Wolff-Fabris, et al., "Anisotropic Dirac Fermions in a Bi Square Net of $SrMnBi_2$," *Physical Review Letters* 107 (2011): 126402, https://doi.org/10.1103/PhysRevLett.107.126402.
27. S. Huan, D. Wang, H. Su, et al., "Magnetism-induced ideal Weyl state in bulk van der Waals crystal $MnSb_2Te_4$," *Applied Physics Letters* 118 (2021): 192105,

https://doi.org/10.1063/5.0047438.
28. B. Cheng, Y. Wang, D. Barbalas, T. Higo, S. Nakatsuji, N. P. Armitage, “Terahertz conductivity of the magnetic Weyl semimetal $Mn_3Sn$ films,” *Applied Physics Letters* 115 (2019): 012405, https://doi.org/10.1063/1.5093414.
29. J. Bai, Q. Dong, B. Ruan, et al., “Large anomalous Hall and Nernst effects in the ferromagnetic semimetal candidate $Mn_3Sn_2$,” *Physical Review B* 109 (2024): 125112, https://doi.org/10.1103/PhysRevB.109.125112.
30. A. B. Childs, S. Baranets, S. Bobev, “Five new ternary indium-arsenides discovered. Synthesis and structural characterization of the Zintl phases $Sr_3In_2As_4$, $Ba_3In_2As_4$, $Eu_3In_2As_4$, $Sr_5In_2As_6$ and $Eu_5In_2As_6$,” *Journal of Solid State Chemistry* 278 (2019): 120889, https://doi.org/10.1016/j.jssc.2019.07.050.
31. S. Balguri, M. B. Mahendru, E. O. G. Delgado, et al., “Two types of colossal magnetoresistance with distinct mechanisms in $Eu_5In_2As_6$,” *Physical Review B* 111 (2025): 115114, https://doi.org/10.1103/PhysRevB.111.115114.
32. Y. Zhang, K. Deng, X. Zhang, et al., “In-plane antiferromagnetic moments and magnetic polaron in the axion topological insulator candidate $EuIn_2As_2$,” *Physical Review B* 101 (2020): 205126, https://doi.org/10.1103/PhysRevB.101.205126.
33. M. S. Cook, E. A. Peterson, C. S. Kengle, et al., “Magnetic polaron formation in $EuZn_2P_2$,” *Physical Review Materials* 9 (2025): 104403, https://doi.org/10.1103/fs97-mpcq.
34. M. V. Ale Crivillero, S. Rößler, S. Granovsky, et al., “Magnetic and electronic properties unveil polaron formation in $Eu_5In_2Sb_6$,” *Scientific Reports* 13 (2023): 1597, https://doi.org/10.1038/s41598-023-28711-z.
35. H. Zhang, F. Du, X. Zheng, et al., “Electronic band reconstruction across the insulator-metal transition in colossally magnetoresistive $EuCd_2P_2$,” *Physical Review B* 108 (2023): L241115, https://doi.org/10.1103/PhysRevB.108.L241115.
36. W. Shon, J.-S. Rhyee, Y. Jin, S.-J. Kim, “Magnetic polaron and unconventional magnetotransport properties of the single-crystalline compound $EuBiTe_3$,” *Physical Review B* 100 (2019): 024433, https://doi.org/10.1103/PhysRevB.100.024433.
37. F. Tang, Y. Chen, W. Yu, et al., “Anisotropic magnetic, magnetotransport, and electronic properties of the layered Zintl compound $EuAl_2Si_2$,” *Physical Review Materials* 9 (2025): 064205, https://doi.org/10.1103/llq5-zddh.
38. M. Pohlit, S. Rößler, Y. Ohno, et al., “Evidence for Ferromagnetic Clusters in the Colossal-Magnetoresistance Material $EuB_6$,” *Physical Review Letters* 120 (2018): 257201, https://doi.org/10.1103/PhysRevLett.120.257201.
39. F. L. A. Machado, P. R. T. Ribeiro, J. Holanda, R. L. Rodríguez-Suárez, A. Azevedo, S. M. Rezende, “Spin-flop transition in the easy-plane antiferromagnet nickel oxide,” *Physical Review B* 95 (2017): 104418, https://doi.org/10.1103/PhysRevB.95.104418.
40. G. J. Nilsen, V. Simonet, C. V. Colin, et al., “Phase diagram of multiferroic $KCu_3As_2O_7(OD)_3$,” *Physical Review B* 95 (2017): 214415, https://doi.org/10.1103/PhysRevB.95.214415.
41. C. L. Huang, K. F. Tseng, C. C. Chou, et al., “Observation of a second metastable spin-ordered state in ferrimagnet $Cu_2OSeO_3$,” *Physical Review B* 83 (2011): 052402, https://doi.org/10.1103/PhysRevB.83.052402.

42. H. Chen, J. Wang, H. Li, et al., "Defect Engineering for Stabilizing Magnetic and Topological Properties in $Mn(Bi_{1-x}Sb_x)_2Te_4$," *Nature Communications* 17 (2026): 1029, https://doi.org/10.1038/s41467-025-67774-6.
43. Y. Shangguan, S. Bao, Z.-Y. Dong, et al., "A one-third magnetization plateau phase as evidence for the Kitaev interaction in a honeycomb-lattice antiferromagnet," *Nature Physics* 19 (2023): 1883–1889, https://doi.org/10.1038/s41567-023-02212-2.
44. J. Tong, J. Parry, Q. Tao, G.-H. Cao, Z.-A. Xu, H. Zeng, "Magnetic properties of EuCuAs single crystal," *Journal of Alloys and Compounds* 602 (2014): 26–31, https://doi.org/10.1016/j.jallcom.2014.02.157.
45. Y. Zhao, Y. Jiang, H. Bae, et al., "Hybrid-order topology in unconventional magnets of Eu-based Zintl compounds with surface-dependent quantum geometry," *Physical Review B* 110 (2024): 205111, https://doi.org/10.1103/PhysRevB.110.205111.
46. M. S. Song, L. Houben, Y. Zhao, et al., "Topotaxial mutual-exchange growth of magnetic Zintl $Eu_3In_2As_4$ nanowires with axion insulator classification," *Nature Nanotechnology* 19 (2024): 1796–1803, https://doi.org/10.1038/s41565-024-01762-7.
47. G. Aivazian, Z. Gong, A. M. Jones, et al., "Magnetic control of valley pseudospin in monolayer $WSe_2$," *Nature Physics* 11 (2015): 148–152, https://doi.org/10.1038/nphys3201.
48. J. Zhang, L. Du, S. Feng, et al., "Enhancing and controlling valley magnetic response in $MoS_2$/$WS_2$ heterostructures by all-optical route," *Nature Communications* 10 (2019): 4226, https://doi.org/10.1038/s41467-019-12128-2.
49. K. L. Seyler, P. Rivera, H. Yu, et al., "Signatures of moiré-trapped valley excitons in $MoSe_2$/$WSe_2$ heterobilayers," *Nature* 567 (2019): 66–70, https://doi.org/10.1038/s41586-019-0957-1.
50. Z. Sun, Z. Cao, J. Cui, et al., "Large Zeeman splitting induced anomalous Hall effect in $ZrTe_5$," *npj Quantum Materials* 5 (2020): 1–7, https://doi.org/10.1038/s41535-020-0239-z.
51. Y.-S. Fu, T. Hanaguri, K. Igarashi, M. Kawamura, M. S. Bahramy, T. Sasagawa, "Observation of Zeeman effect in topological surface state with distinct material dependence," *Nature Communications* 7 (2016): 10829, https://doi.org/10.1038/ncomms10829.
52. Y. Liu, X. Yuan, C. Zhang, et al., "Zeeman splitting and dynamical mass generation in Dirac semimetal $ZrTe_5$," *Nature Communications* 7 (2016): 12516, https://doi.org/10.1038/ncomms12516.
53. D. Santos-Cottin, I. Mohelský, J. Wyzula, et al., "$EuCd_2As_2$: A Magnetic Semiconductor," *Physical Review Letters* 131 (2023): 186704, https://doi.org/10.1103/PhysRevLett.131.186704.
54. S. Nasrallah, D. Santos-Cottin, F. Le Mardelé, et al., "Magneto-optical response of the magnetic semiconductors $EuCd_2X_2$ (X=P, As, Sb)," *Physical Review B* 110 (2024): L201201, https://doi.org/10.1103/PhysRevB.110.L201201.
55. W. Wu, Z. Shi, M. Ozerov, et al., "The discovery of three-dimensional Van Hove singularity," *Nature Communications* 15 (2024): 2313, https://doi.org/10.1038/s41467-024-46626-9.
56. L. E. Schmutz, G. Dresselhaus, M. S. Dresselhaus, "Optical absorption of EuTe in high magnetic fields," *Solid State Communications* 28 (1978): 597–600, https://doi.org/10.1016/0038-1098(78)90588-4.
57. J. S. Moodera, T. S. Santos, T. Nagahama, "The phenomena of spin-filter tunnelling," *Journal of Physics: Condensed Matter* 19 (2007): 165202, https://doi.org/10.1088/0953-

8984/19/16/165202.
58. L. L. Hirst, “Theory of the coupling between conduction electrons and moments of 3d and 4f ions in metals,” *Advances in Physics* 27 (1978): 231–285, https://doi.org/10.1080/00018737800101374.
59. R. Yu, W. Zhang, H.-J. Zhang, S.-C. Zhang, X. Dai, Z. Fang, “Quantized Anomalous Hall Effect in Magnetic Topological Insulators,” *Science* 329 (2010): 61–64, https://doi.org/10.1126/science.1187485.
60. Y. Zhao, H. Liu, J. Yan, et al., “Anisotropic magnetotransport and exotic longitudinal linear magnetoresistance in $WTe_2$ crystals,” *Physical Review B* 92 (2015): 041104, https://doi.org/10.1103/PhysRevB.92.041104.
61. N. Kumar, Y. Sun, N. Xu, et al., “Extremely high magnetoresistance and conductivity in the type-II Weyl semimetals $WP_2$ and $MoP_2$,” *Nature Communications* 8 (2017): 1642, https://doi.org/10.1038/s41467-017-01758-z.
62. X. Sun, F. Tang, X. Shen, et al., “Anisotropic giant magnetoresistance and Fermi surface topology in the layered compound $YbBi_2$,” *Physical Review B* 105 (2022): 195114, https://doi.org/10.1103/PhysRevB.105.195114.
63. Q. Dong, P. Yang, Z. Liu, et al., “Simultaneous colossal magnetoresistance and angular magnetoresistance in the antiferromagnetic semiconductor $EuSe_2$,” *Physical Review B* 112 (2025): L140405, https://doi.org/10.1103/p2c5-r163.
64. H. Yang, Q. Liu, Z. Liao, et al., “Colossal angular magnetoresistance in the antiferromagnetic semiconductor $EuTe_2$,” *Physical Review B* 104 (2021): 214419, https://doi.org/10.1103/PhysRevB.104.214419.
65. Z. L. Sun, A. F. Wang, H. M. Mu, et al., “Field-induced metal-to-insulator transition and colossal anisotropic magnetoresistance in a nearly Dirac material $EuMnSb_2$,” *npj Quantum Materials* 6 (2021): 94, https://doi.org/10.1038/s41535-021-00397-4.
66. M. Lyu, J. Xiang, Z. Mi, et al., “Nonsaturating magnetoresistance, anomalous Hall effect, and magnetic quantum oscillations in the ferromagnetic semimetal PrAlSi,” *Physical Review B* 102 (2020): 085143, https://doi.org/10.1103/PhysRevB.102.085143.
67. J. Klier, I. V. Gornyi, A. D. Mirlin, “Transversal magnetoresistance and Shubnikov–de Haas oscillations in Weyl semimetals,” *Physical Review B* 96 (2017): 214209, https://doi.org/10.1103/PhysRevB.96.214209.
68. T. Liang, Q. Gibson, M. N. Ali, M. Liu, R. J. Cava, N. P. Ong, “Ultrahigh mobility and giant magnetoresistance in the Dirac semimetal $Cd_3As_2$,” *Nature Materials* 14 (2015): 280–284, https://doi.org/10.1038/nmat4143.
69. S. Wang, B.-C. Lin, A.-Q. Wang, D.-P. Yu, Z.-M. Liao, “Quantum transport in Dirac and Weyl semimetals: a review,” *Advances in Physics: X* 2 (2017): 518–544, https://doi.org/10.1080/23746149.2017.1327329.
70. C. Shekhar, A. K. Nayak, Y. Sun, et al., “Extremely large magnetoresistance and ultrahigh mobility in the topological Weyl semimetal candidate NbP,” *Nature Physics* 11 (2015): 645–649, https://doi.org/10.1038/nphys3372.
71. M. N. Ali, J. Xiong, S. Flynn, et al., “Large, non-saturating magnetoresistance in $WTe_2$,” *Nature* 514 (2014): 205–208, https://doi.org/10.1038/nature13763.
72. Q. Jiang, J. C. Palmstrom, J. Singleton, et al., “Revealing Fermi surface evolution and Berry curvature in an ideal type-II Weyl semimetal,” *Nature Communications* 15 (2024): 2310,

https://doi.org/10.1038/s41467-024-46633-w.
73. S. H. Lee, D. Graf, L. Min, et al., "Evidence for a Magnetic-Field-Induced Ideal Type-II Weyl State in Antiferromagnetic Topological Insulator $Mn(Bi_{1-x}Sb_x)_2Te_4$," *Physical Review X* 11 (2021): 031032, https://doi.org/10.1103/PhysRevX.11.031032.
74. S. Mugiraneza, A. M. Hallas, "Tutorial: a beginner's guide to interpreting magnetic susceptibility data with the Curie-Weiss law," *Communications Physics* 5 (2022): 95, https://doi.org/10.1038/s42005-022-00853-y.
75. W. Wu, Z. Shi, Y. Du, et al., "Topological Lifshitz transition and one-dimensional Weyl mode in $HfTe_5$," *Nature Materials* 22 (2023): 84–91, https://doi.org/10.1038/s41563-022-01364-5.
76. P. Santini, S. Carretta, G. Amoretti, R. Caciuffo, N. Magnani, G. H. Lander, "Multipolar interactions in *f*-electron systems: The paradigm of actinide dioxides," *Reviews of Modern Physics* 81 (2009): 807–863, https://doi.org/10.1103/RevModPhys.81.807.
77. E. Cheng, K. Wang, Y. Hao, et al., "Interwoven magnetic kagome metal overcomes geometric frustration," *Nature Materials* 25 (2026): 602–609, https://doi.org/10.1038/s41563-025-02414-4.
78. K. Guo, Z. Ma, H. Liu, et al., "1/3 and other magnetization plateaus in the quasi-one-dimensional Ising magnet $TbTi_3Bi_4$ with zigzag spin chain," *Physical Review B* 110 (2024): 064416, https://doi.org/10.1103/PhysRevB.110.064416.
79. A. V. Savinkov, S. L. Korableva, A. A. Rodionov, et al., "Magnetic properties of $Dy^{3+}$ ions and crystal field characterization in $YF_3$:$Dy^{3+}$ and $DyF_3$ single crystals," *Journal of Physics: Condensed Matter* 20 (2008): 485220, https://doi.org/10.1088/0953-8984/20/48/485220.
80. Y. Lai, L. Ke, J. Yan, R. D. McDonald, R. J. McQueeney, "Defect-driven ferrimagnetism and hidden magnetization in $MnBi_2Te_4$," *Physical Review B* 103 (2021): 184429, https://doi.org/10.1103/PhysRevB.103.184429.
81. J.-Q. Yan, S. Okamoto, M. A. McGuire, A. F. May, R. J. McQueeney, B. C. Sales, "Evolution of structural, magnetic, and transport properties in $MnBi_{2-x}Sb_xTe_4$," *Physical Review B* 100 (2019): 104409, https://doi.org/10.1103/PhysRevB.100.104409.
82. M. M. Otrokov, I. I. Klimovskikh, H. Bentmann, et al., "Prediction and observation of an antiferromagnetic topological insulator," *Nature* 576 (2019): 416–422, https://doi.org/10.1038/s41586-019-1840-9.
83. B.-X. Li, Z. Song, Z. Fang, Z. Wang, H. Weng, "Manipulation of topological phase transitions and the mechanism of magnetic interactions in Eu-based Zintl-phase materials," *Physical Review B* 111 (2025): 205127, https://doi.org/10.1103/PhysRevB.111.205127.
84. J. Shuai, J. Mao, S. Song, Q. Zhang, G. Chen, Z. Ren, "Recent progress and future challenges on thermoelectric Zintl materials," *Materials Today Physics* 1 (2017): 74–95, https://doi.org/10.1016/j.mtphys.2017.06.003.
85. R. Nesper, "Structure and chemical bonding in zintl-phases containing lithium," *Progress in Solid State Chemistry* 20 (1990): 1–45, https://doi.org/10.1016/0079-6786(90)90006-2.
86. J. Jiang, M. M. Olmstead, S. M. Kauzlarich, H.-O. Lee, P. Klavins, Z. Fisk, "Negative Magnetoresistance in a Magnetic Semiconducting Zintl Phase: $Eu_3In_2P_4$," *Inorganic Chemistry* 44 (2005): 5322–5327, https://doi.org/10.1021/ic0504036.
87. J. Li, Y. Li, S. Du, et al., "Intrinsic magnetic topological insulators in van der Waals layered $MnBi_2Te_4$-family materials," *Science Advances* 5 (2019): eaaw5685,

https://doi.org/10.1126/sciadv.aaw5685.

88. J.-R. Soh, F. De Juan, M. G. Vergniory, et al., “Ideal Weyl semimetal induced by magnetic exchange,” *Physical Review B* 100 (2019): 201102, https://doi.org/10.1103/PhysRevB.100.201102.

89. S. Nie, T. Hashimoto, F. B. Prinz, “Magnetic Weyl Semimetal in $K_2Mn_3(AsO_4)_3$ with the Minimum Number of Weyl Points,” *Phys. Rev. Lett* 128 (2022): 176401, https://doi.org/10.1103/PhysRevLett.128.176401.

90. I. Belopolski, R. Watanabe, Y. Sato, et al., “Synthesis of a semimetallic Weyl ferromagnet with point Fermi surface,” *Nature* 637 (2025): 1078–1083, https://doi.org/10.1038/s41586-024-08330-y.

91. Y. Wang, J. Ma, J. Yuan, et al., “Absence of metallicity and bias-dependent resistivity in low-carrier-density $EuCd_2As_2$,” *Science China Physics, Mechanics & Astronomy* 67 (2024): 247311, https://doi.org/10.1007/s11433-023-2283-0.

92. Y. Shi, Z. Liu, L. A. Burnett, et al., “Absence of Weyl nodes in $EuCd_2As_2$ revealed by the carrier density dependence of the anomalous Hall effect,” *Physical Review B* 109 (2024): 125202, https://doi.org/10.1103/PhysRevB.109.125202.

93. K. M. Taddei, K. G. S. Ranmohotti, D. S. Liurukara, et al., “Insulating ground state and 2-*k* magnetic structure of candidate Weyl Hydrogen atom $K_2Mn_3(AsO_4)_3$,” *Physical Review B* 113 (2026): 024423, https://doi.org/10.1103/9j29-yb8p.

94. K. Jia, J. Yao, X. He, et al., “Discovery of a Magnetic Topological Semimetal $Eu_3In_2As_4$ with a Single Pair of Weyl Points,” *arXiv*, (2024), https://doi.org/10.48550/arXiv.2403.07637.

95. Z. Shi, W. Wu, Z. Zhang, et al., “A high-flux and high-efficiency setup for magneto-infrared spectroscopy,” *Review of Scientific Instruments* 96 (2025): 113902, https://doi.org/10.1063/5.0296925.

96. P. E. Blöchl, “Projector augmented-wave method,” *Physical Review B* 50 (1994): 17953–17979, https://doi.org/10.1103/PhysRevB.50.17953.

97. G. Kresse, J. Furthmüller, “Efficient iterative schemes for ab initio total-energy calculations using a plane-wave basis set,” *Physical Review B* 54 (1996): 11169–11186, https://doi.org/10.1103/PhysRevB.54.11169.

98. J. P. Perdew, K. Burke, M. Ernzerhof, “Generalized Gradient Approximation Made Simple,” *Physical Review Letters* 77 (1996): 3865–3868, https://doi.org/10.1103/PhysRevLett.77.3865.

99. S. L. Dudarev, G. A. Botton, S. Y. Savrasov, C. J. Humphreys, A. P. Sutton, “Electron-energy-loss spectra and the structural stability of nickel oxide: An LSDA+U study,” *Physical Review B* 57 (1998): 1505–1509, https://doi.org/10.1103/PhysRevB.57.1505.

100. A. Togo, I. Tanaka, “First principles phonon calculations in materials science,” *Scripta Materialia* 108 (2015): 1–5, https://doi.org/10.1016/j.scriptamat.2015.07.021.

101. I. Souza, N. Marzari, D. Vanderbilt, “Maximally localized Wannier functions for entangled energy bands,” *Physical Review B* 65 (2001): 035109, https://doi.org/10.1103/PhysRevB.65.035109.

102. Q. Wu, S. Zhang, H.-F. Song, M. Troyer, A. A. Soluyanov, “WannierTools: An open-source software package for novel topological materials,” *Computer Physics Communications* 224 (2018): 405–416, https://doi.org/10.1016/j.cpc.2017.09.033.

103. J. Wyzula, I. Mohelský, D. Václavková, et al., "High-Angular Momentum Excitations in Collinear Antiferromagnet $FePS_3$," *Nano Letters* 22 (2022): 9741–9747, https://doi.org/10.1021/acs.nanolett.2c04111.
104. P. Chen, B. S. Holinsworth, K. R. O'Neal, et al., "Magnetic-field-induced shift of the optical band gap in $Ni_3V_2O_8$," *Physical Review B* 89 (2014): 165120, https://doi.org/10.1103/PhysRevB.89.165120.
105. J. Klein, Z. Song, B. Pingault, et al., "Sensing the Local Magnetic Environment through Optically Active Defects in a Layered Magnetic Semiconductor," *ACS Nano* 17 (2023): 288–299, https://doi.org/10.1021/acsnano.2c07655.
106. F. Le Mardelé, J. Wyzula, I. Mohelsky, et al., "Evidence for three-dimensional Dirac conical bands in TlBiSSe by optical and magneto-optical spectroscopy," *Physical Review B* 107 (2023): L241101, https://doi.org/10.1103/PhysRevB.107.L241101.
107. J.-X. Yin, W. Ma, T. A. Cochran, et al., "Quantum-limit Chern topological magnetism in $TbMn_6Sn_6$," *Nature* 583 (2020): 533–536, https://doi.org/10.1038/s41586-020-2482-7.
108. Z.-G. Chen, R. Y. Chen, R. D. Zhong, et al., "Spectroscopic evidence for bulk-band inversion and three-dimensional massive Dirac fermions in $ZrTe_5$," *Proceedings of the National Academy of Sciences* 114 (2017): 816–821, https://doi.org/10.1073/pnas.1613110114.

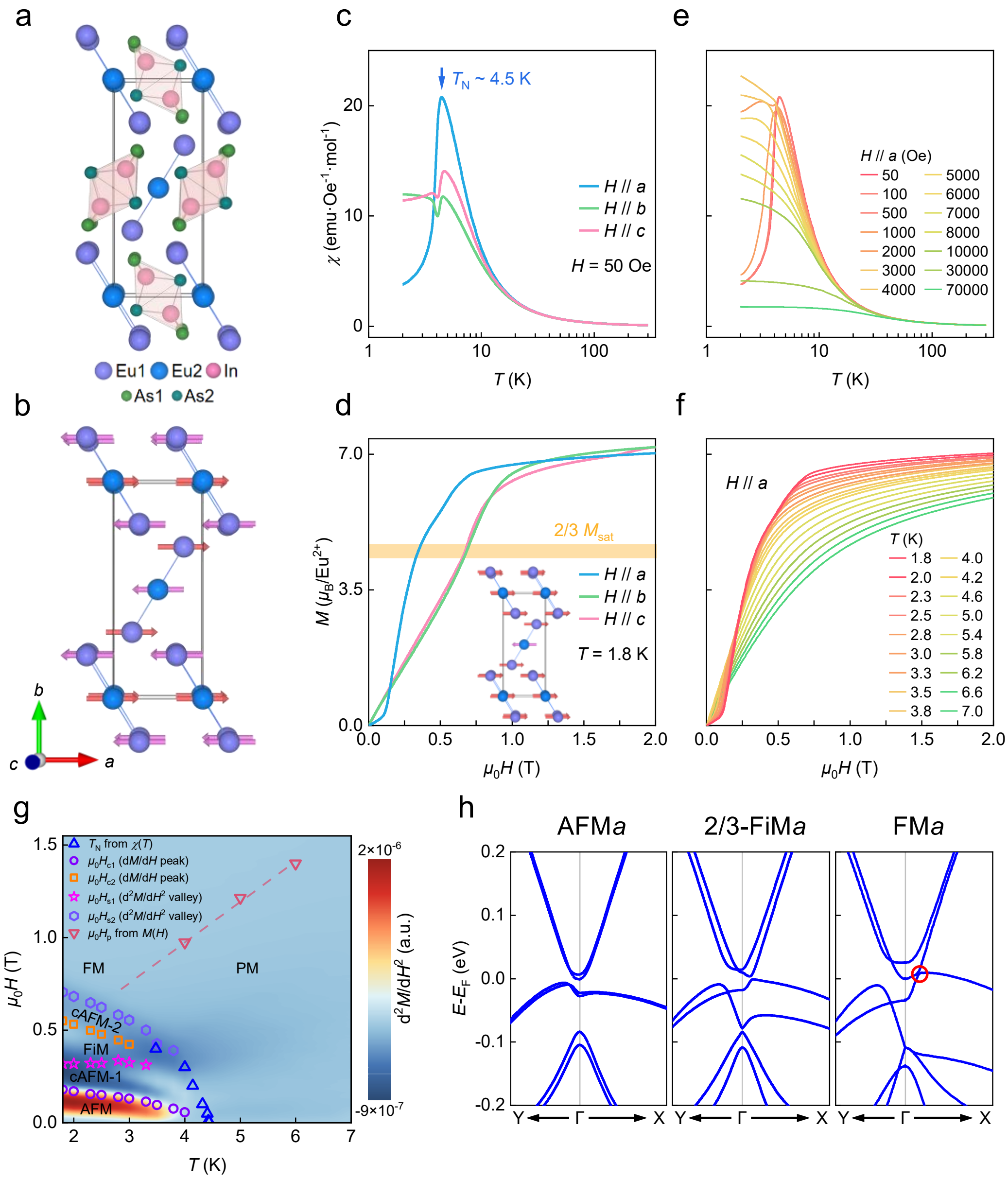

**FIGURE 1. Crystal structure and magnetism of $Eu_3In_2As_4$.**

**a** Schematic crystal structure of $Eu_3In_2As_4$. The inequivalent Eu sublattices are represented by purple and blue atoms

**b** Magnetic structure of the C-type AFM*a* ground state, forming FM coupling along *c*-axis and AFM coupling within the *ab*-plane. For clarity, only Eu atoms are shown here. Colored arrows denote oppositely oriented spins, with the two Eu sublattices each forming AFM structures.

**c** Temperature dependence of magnetic susceptibility measured under field-cooled (FC) conditions with an applied field of 50 Oe along the three principal crystallographic axes.

**d** Field-dependent magnetization at 1.8 K with the field applied along three crystallographic axes. The yellow shaded region denotes the 2/3 $M_{sat}$ regime, with its upper and lower boundaries defined by two-thirds of the magnetization measured at 7 T and at the ferromagnetic saturation field, respectively.

**e** Temperature-dependent susceptibility measured under increasing magnetic field along *a*-axis.

**f** Field-dependent magnetization at various temperatures with the field applied along *a*-axis.

**g** Magnetic phase diagram of $Eu_3In_2As_4$ for $H$ // $a$. The color plot background exhibits the evolution of $d^2M/dH^2$ with field and temperature. Features in the second derivative evidence the 2/3-FiM$a$ state. The red dashed line is a guide to the eye denoting the FM–PM phase boundary.

**h** Band structures of the AFM$a$, 2/3-FiM$a$, and FM$a$ states (from left to right, respectively) along Y–Γ–X path, illustrating a continuous bandgap closing process.

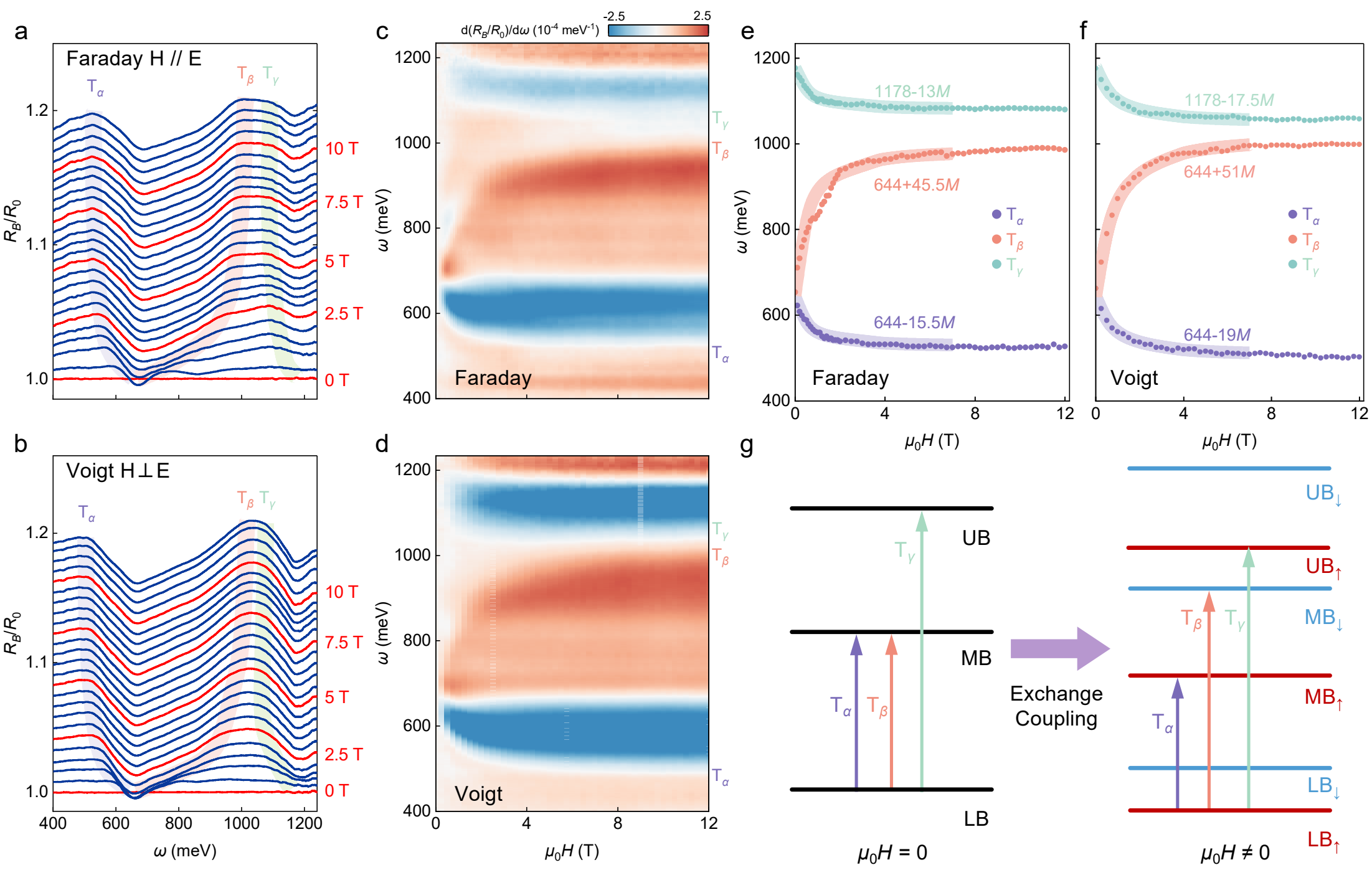

**FIGURE 2. Magneto-infrared spectra of $Eu_3In_2As_4$.**

**a, b** Stacked plots of relative magneto-reflectivity $R_B/R_0$ spectra (0.5 T step) measured in Faraday ($H$ // $a$) and Voigt ($H$ // $c$) geometries, respectively, with the infrared beam focused on the (100) plane. Three distinct sets of peak features, denoted $T_\alpha$, $T_\beta$, and $T_\gamma$, appear and evolve with the increasing field, and are respectively traced by violet, red and green curves.

**c, d** False-color plots of the first derivative of relative magneto-reflectivity $d(R_B/R_0)/d\omega$, in (**c**) Faraday and (**d**) Voigt configurations, respectively.

**e, f** Correlation between magnetization and optical transition energies with Faraday and Voigt configurations in (**e**) and (**f**), respectively. Colored dots denote the apparent energies of $T_\alpha$, $T_\beta$, and $T_\gamma$. Shaded curves indicate the linear scaling of magnetization curves around 6 K, with scaling parameters labeled near each curve.

**g** Possible origin of optical transitions. Solid arrows represent the optical transitions predicted by the model. Black lines indicate the spin-degenerate bands, while red and blue distinguish spin-polarized bands under exchange-induced splitting.

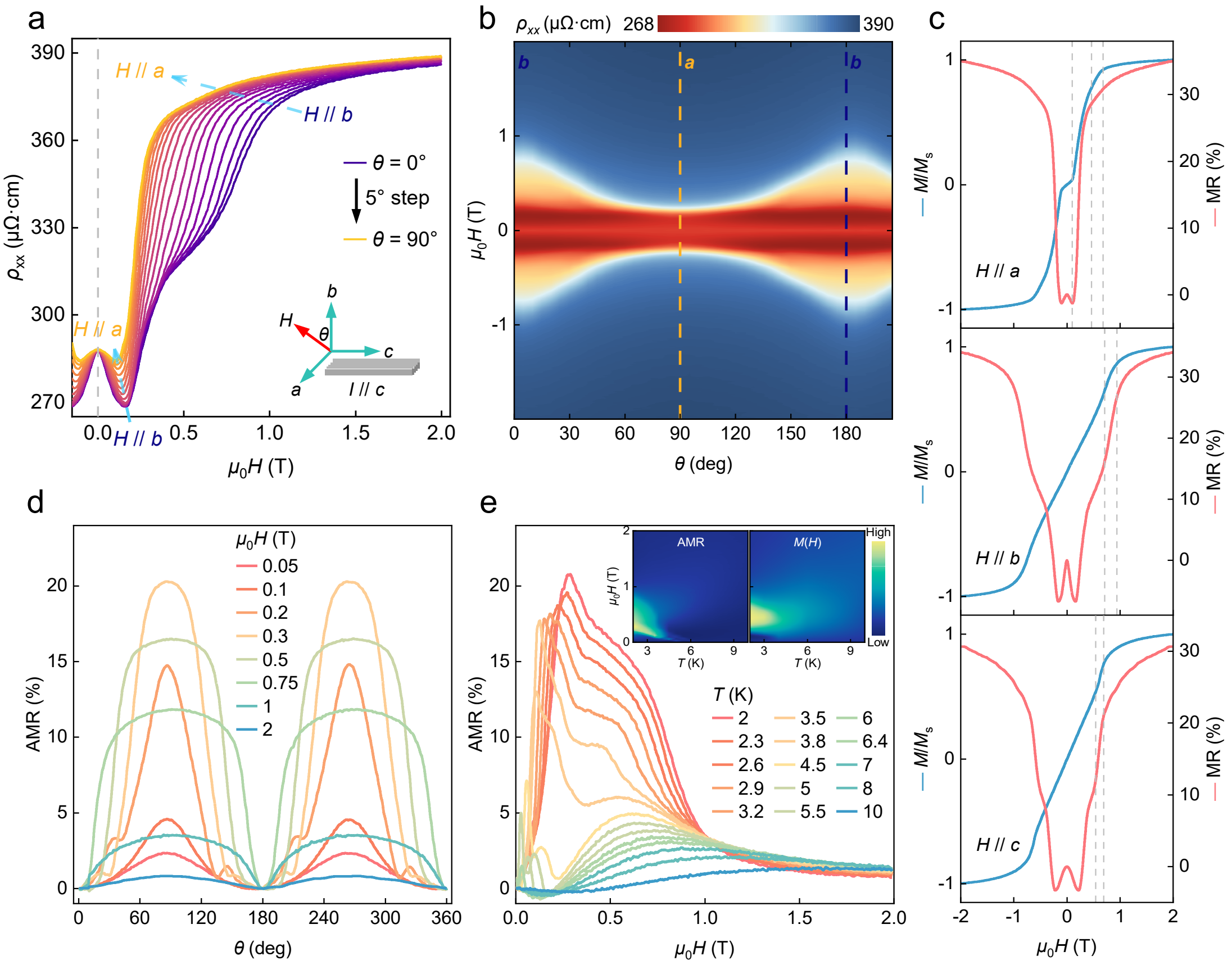

**FIGURE 3. Anisotropic magneto-transport in the low-field regime.**

**a** Field dependence of the magneto-resistivity, $\rho_{xx}(H)$, at 2 K. Current was applied along the *c*-axis while the field direction was rotated from the *b*- to *a*-axis by angle $\theta$, as illustrated in the inset.

**b** Color plot of the angular-dependent $\rho_{xx}(\theta, H)$, revealing *ab*-plane anisotropy.

**c** Comparison between the magneto-resistance ($MR = \frac{[\rho_{xx}(H) - \rho_{xx}(0\ \mathrm{T})]}{\rho_{xx}(0\ \mathrm{T})} \times 100\%$) and magnetization curves (normalized by the saturation magnetization $M_s$) at 2 K for field along the three primary axes. Anomalies in the magnetization can be reflected by features in the MR curves.

**d** Angular-dependent MR (AMR) at selected fields (2 K), with the field rotated from 0° to 360° within the *ab*-plane. The AMR shows a two-fold symmetry with maxima at 90° (*H* // *a*).

**e** Field dependence of AMR at 90° at increasing temperatures. The AMR shows a non-monotonic field dependence, peaking around 0.3 T and decreasing towards zero at higher fields. This non-monotonic behavior gradually diminishes as temperature increases above $T_N$. The insets show the color plot of the AMR($H$, $T$) at 90° and magnetization difference between *a*- and *b*- axis ($M_a(H, T)$-$M_b(H, T)$), revealing a similar trend.

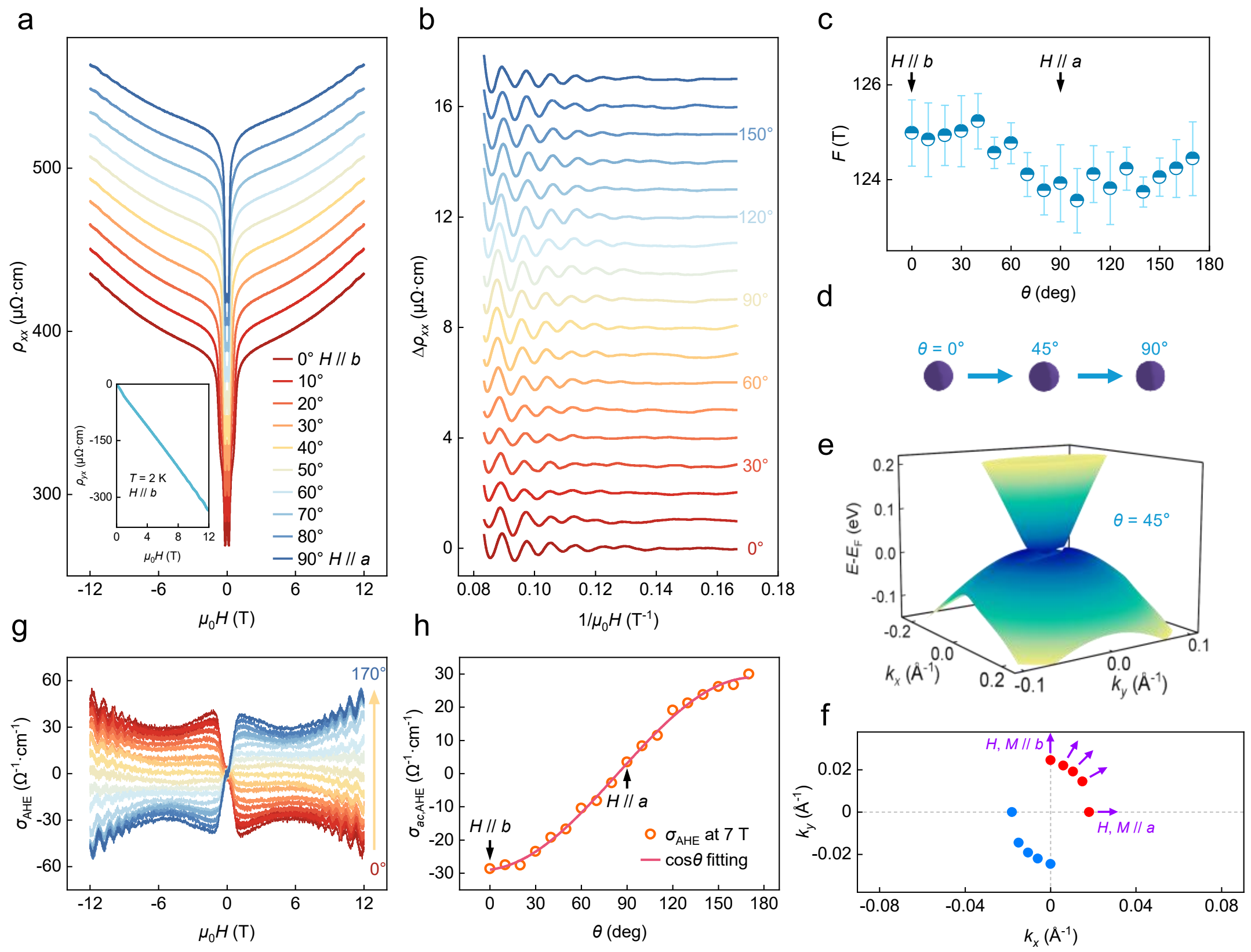

**FIGURE 4. Shubnikov-de Haas (SdH) oscillations and anomalous Hall effect (AHE) in FM-$Eu_3In_2As_4$.**

**a** Magneto-resistivity $\rho_{xx}(H)$ at 2 K within ±12 T, with $H$ rotated within *ab*-plane. Current was applied along *c*-axis. The $\rho_{xx}(H)$ curves are offset for clarity. Inset: Hall resistivity $\rho_{yx}(H)$ at 0°, exhibiting a nearly linear field dependence with negative slope, indicating electron conduction. Linear fitting of $\rho_{yx}(H)$ in FM state yields an electron density of $2.1\times10^{19}$ cm$^{-3}$.

**b** SdH oscillations $\Delta\rho_{xx}$ (obtained by polynomial fitting) as a function of $1/\mu_0H$ with varying field directions.

**c** SdH oscillation frequencies (with error bars) versus field direction $\theta$, showing weak angular dependence. The frequency was extracted as linear-fit slope of the Landau index $n$ versus $1/\mu_0H$ in Landau fan diagrams.

**d** Calculated 3D Fermi surfaces of $Eu_3In_2As_4$ for three representative magnetization directions ($\theta$ = 0°, 45°, and 90°). For each configuration, the viewing direction is aligned with the applied magnetic field, so that the projected cross section corresponds to the extremal orbit perpendicular to $H$. The Fermi surface is plotted at $E$-$E_F$ = 0.155 eV, chosen to match the experimentally extracted SdH frequency.

**e** 3D band structure of FM $Eu_3In_2As_4$ with $\theta$ = 45°, demonstrating a single pair of Weyl points located in the $k_x$-$k_y$ plane. For clarity, only the two bands forming the Weyl crossing are shown.

**f** Position evolution of single Weyl pairs of FM $Eu_3In_2As_4$, with the field (magnetization) is rotated in the *ab*-plane ($\theta$ = 0°, 30°, 45°, 60°, 90°), denoted as the purple arrows. Blue and red colors represent opposite chirality of Weyl points.

**g** Field-dependent anomalous Hall conductivity $\sigma_{AHE}$ at 2 K with rotating $H$, where $\sigma_{AHE} = \rho_{AHE}/(\rho_{xx}^2 + \rho_{yx}^2)$ and $\rho_{AHE} = \rho_{yx} - R_H\mu_0H$ ($R_H$ is the linear Hall coefficient in the FM state).

**h** Angular-dependent $\sigma_{\mathrm{AHE}}$ values (the $\sigma_{ac,\mathrm{AHE}}$ component) at 7 T, exhibiting a cos$\theta$ trend.

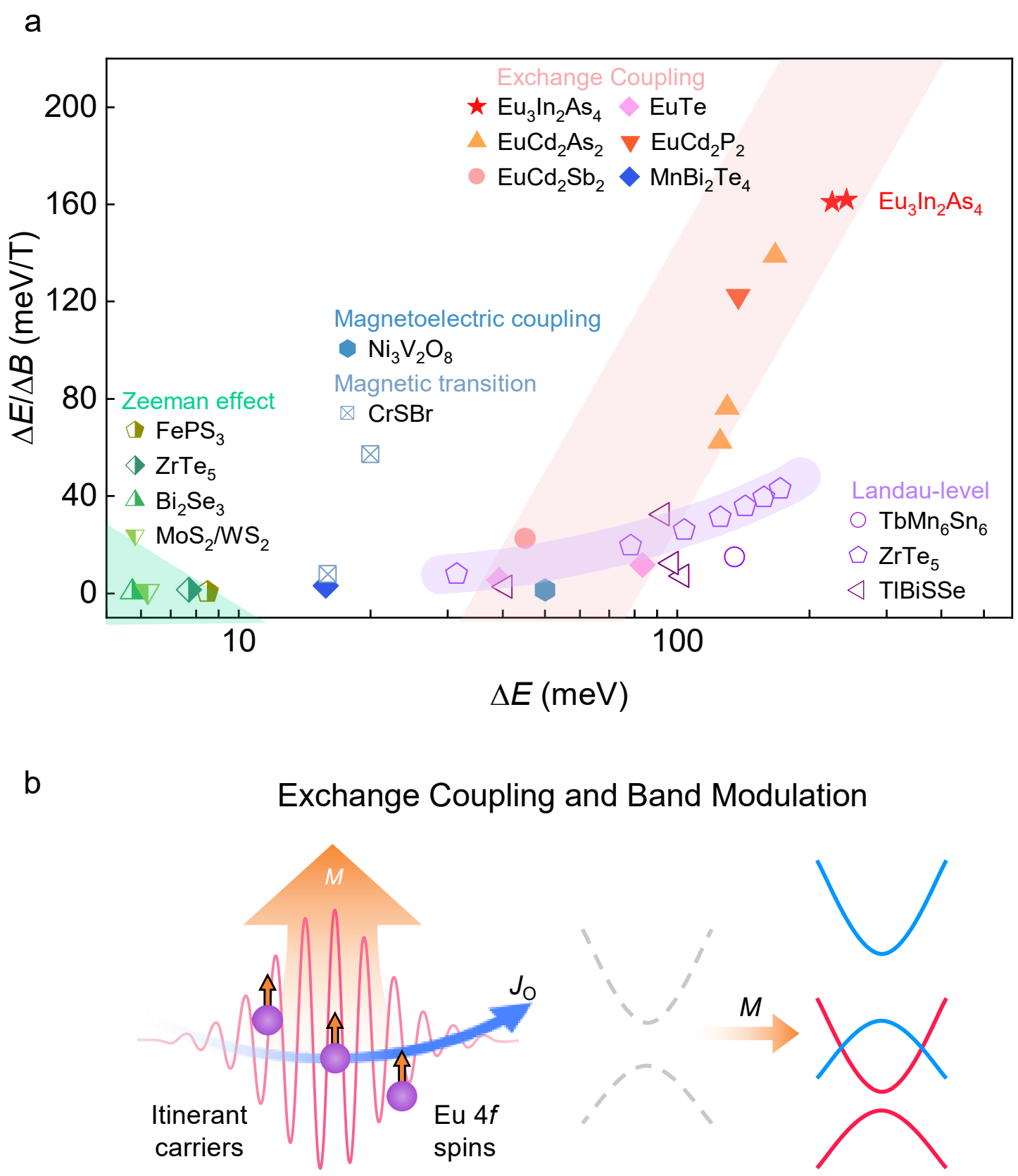

**FIGURE 5. The superiority of Eu$^{2+}$-based exchange coupling in comparison with other mechanisms and its principle schematic.**

**a** Comparison of energy scales and effective average field slopes ($\Delta E/\Delta B$) across various modulation mechanisms: exchange coupling[23, 53–56], Zeeman effect[48, 50, 51, 103], magnetoelectric coupling[104], magnetic order transition[105], and Landau quantization[106–108].

**b** Illustration of the exchange coupling between localized spins and itinerant carriers. Purple balls refer to Eu$^{2+}$ ions, and the electrons are illustrated as wave packet. The right panel demonstrates the M-induced exchange splitting and crossing, promoting the formation of Weyl points.